\documentclass[prc,aps,twocolumn,floats,floatfix,showpacs,superscriptaddress,nofootinbib]{revtex4}
\usepackage{amsmath}
\usepackage{amsfonts}
\usepackage{graphicx}

\begin{document}

\title{Influence of 2p-2h configurations on $\beta$-decay rates}

\author{A. P. Severyukhin}
\affiliation{Bogoliubov Laboratory of Theoretical Physics,
             Joint Institute for Nuclear Research,
             141980 Dubna, Moscow region, Russia}
\author{V. V. Voronov}
\affiliation{Bogoliubov Laboratory of Theoretical Physics,
             Joint Institute for Nuclear Research,
             141980 Dubna, Moscow region, Russia}
\author{I. N. Borzov}
\affiliation{Bogoliubov Laboratory of Theoretical Physics,
             Joint Institute for Nuclear Research,
             141980 Dubna, Moscow region, Russia}
\author{N. N. Arsenyev}
\affiliation{Bogoliubov Laboratory of Theoretical Physics,
             Joint Institute for Nuclear Research,
             141980 Dubna, Moscow region, Russia}
\author{Nguyen Van Giai}
\affiliation{Institut de Physique Nucl\'eaire, CNRS-IN2P3 and
             Univ. Paris-Sud, 91405 Orsay, France}

\begin{abstract}
The effects of the phonon-phonon coupling on the $\beta$-decay rates of neutron-rich
nuclei are studied in a microscopic model based on Skyrme-type interactions. The approach
uses a finite-rank separable approximation (FRSA) of the Skyrme-type particle-hole (p-h)
residual interaction. Very large two-quasiparticle spaces can thus be treated.
A redistribution of the Gamow-Teller (G-T) strength is found due to the tensor correlations
and the $2p-2h$ fragmentation of G-T states. As a result, the $\beta^-$-decay half-lives are
decreased significantly. Using the Skyrme interaction SGII together with a volume-type
pairing interaction we illustrate this reduction effect by comparing with available experimental
data for the Ni isotopes and  neutron-rich $N=50$ isotones.
We give predictions for $^{76}$Fe and $^{80}$Ni in comparison with the case
of the doubly-magic nucleus $^{78}$Ni which is an important waiting point
in the r-process.
\end{abstract}


\date{June 26, 2014}

\maketitle
%
%=======================================================
%
\section{Introduction}
Many fundamental nuclear physics issues depend on our quantitative
understanding of the $\beta$-decay phenomena in nuclei. Due to phase-space
amplification effects, the $\beta$-decay rates are sensitive to both
nuclear binding energies and $\beta$-strength functions.
Within an appropriate $\beta$-decay model, the correct amount of the
integral $\beta$-strength should be placed within the properly calculated
$Q_{\beta}$- window   provided that the  spectral distribution is
also close to the "true" $\beta$-strength function.
It is desirable to have theoretical models which can describe the
data wherever they can be measured, and predict the properties related
to spin-isospin modes in the nuclei too short-lived to allow for
experimental studies. One of the successful tools for studying
charge-exchange nuclear modes is the quasiparticle random phase
approximation (QRPA) with the self-consistent mean-field derived
from a Skyrme-type energy-density functional (EDF), see e.g.,
\cite{inbsf96,ebnds99,inbsg00,bden02,colo07,bszzcx09,sag11,mb13}.
These QRPA calculations enable one to describe the properties of
the ground state and excited charge-exchange states using the same
EDF.

Experimental studies using the multipole decomposition analysis of
the (n,p) and (p,n) reactions~\cite{w97,mda} found substantial
Gamow-Teller (G-T) strength above the G-T resonance peak and have
clarified a longstanding problem of the missing experimental G-T
strength, hence resolving the discrepancies between the theoretical
RPA predictions and the experimental measurements.
It has been found necessary to take into account the
coupling with more complex configurations in order to shift
some strength to higher energies and to comply with the experimental results
\cite{bh82,ks84,dosw87}. Using the Skyrme EDF and the RPA, such
attempts in the past~\cite{cnbb94,csnbs98} have allowed one to
understand the damping of charge-exchange resonances and their
particle decay. Recently, the damping of the G-T mode has been
investigated using the Skyrme-RPA plus particle-vibration
coupling~\cite{ncbbm12}. The main difficulty is that the complexity
of the calculations increases rapidly with the size of the
configuration space and one has to work within limited spaces.

Making use of the finite rank separable approximation (FRSA)
\cite{gsv98,ssvg02,svg08} for the residual interaction one can
perform Skyrme-QRPA calculations in very large two-quasiparticle
spaces. Following the basic ideas of the quasiparticle-phonon
model (QPM)~\cite{solo}, the approach has been generalized for the
coupling between one- and two-phonon components of the wave
functions~\cite{svg04}. The so-called FRSA was thus used to study
the electric low-lying states and giant resonances within and
beyond the QRPA ~\cite{svg08,svg04,sap12}.

Recently, the FRSA approach was extended to charge-exchange nuclear
excitations~\cite{svg12} and also for accommodating the  tensor
correlations which mimic the Skyrme-type tensor interactions~\cite{ss13}.
In the present work we generalize the approach to the coupling
between one- and two-phonon components in the wave functions.
As an application of the method we study the $\beta$-decay half-lives
of neutron-rich $N=50$ isotones and Ni isotopes and we compare
to the most neutron-rich ($(N-Z)/A=0.28$) doubly-magic nucleus
$^{78}$Ni which is also an important waiting point in the
r-process~\cite{78ni}. In the case of $^{78}$Ni preliminary results
of our calculation without the tensor interaction
are already reported in Ref.~\cite{svbg13}.

This paper is organized as follows. In Sec.~II, we sketch the
method for including the effects of the phonon-phonon coupling.
In Sec. III, we discuss the details of QRPA calculations for
the $1^+$ states of the daughter nuclei and the $2^+$ states of
the parent nuclei. In Sec.~IV, we analyze the results of the
calculations of $\beta$-decay rates. Conclusions are
finally drawn in Sec.~V.
%
%===============================================================
%
\section{The FRSA model}
The FRSA model for charge-exchange excitations was already
introduced in Refs.~\cite{svg12,ss13}.
In the present study,
this method is extended by including the coupling between one- and
two-phonon terms in the wave functions of G-T states.
The starting point
is the Hartree-Fock(HF)-BCS calculation~\cite{RingSchuck} of the parent ground state within
a spherical symmetry assumption.
In the particle-hole (p-h) channel we use the Skyrme
interaction with the triplet-even and triplet-odd tensor components introduced in Refs.~\cite{s59,sbf77}.
The continuous part of the
single-particle spectrum is discretized by diagonalizing the HF
hamiltonian on a harmonic oscillator basis. The inclusion of the
tensor interaction results in the following modification of the
spin-orbit potential in coordinate space~\cite{hsplb07,t44}:
\begin{equation}
\label{sop}
U^{(q)}_{S.O.}=\frac{W_0}{2r}\left(2\frac{d\rho_q}{dr}+\frac{d\rho_{q'}}{dr}\right)
+\left(\alpha\frac{J_q}{r}+\beta\frac{J_{q'}}{r}\right),
\end{equation}
where $\rho_q$ and $J_q$ ($q=n,p$) are the densities and the
spin-orbit densities, respectively. The coefficients $\alpha$ and
$\beta$ can be separated into contributions from the central force
($\alpha_c$, $\beta_c$) and the tensor force ($\alpha_T$,
$\beta_T$)~\cite{hsplb07,t44}. The pairing correlations are
generated by the density-dependent zero-range force
\begin{equation}
\label{pair} V_{pair}({\bf r}_1,{\bf r}_2)=V_{0}\left(
1-\eta\left(\frac{\rho \left( r_{1}\right) } {\rho
_{0}}\right)^{\gamma}\right) \delta \left( {\bf r}_{1}-{\bf
r}_{2}\right),
\end{equation}
where
$\rho _{0}$ is the nuclear saturation density.
The values of $V_{0}$, $\eta$ and $\gamma$ are fixed to reproduce the
odd-even mass difference of the studied nuclei~\cite{svg08}.

To build the QRPA equations on the basis of HF-BCS quasiparticle states with
the residual interactions consistently derived from the Skyrme EDF in the p-h channel and from
the zero-range pairing force in the particle-particle (p-p) channel is
a standard procedure~\cite{t05}. The dimensions of the QRPA matrix
grow very rapidly with the size of the nuclear system unless severe
and damaging cut-offs are made to the 2-quasiparticle configuration space.
It is well known that, if the QRPA matrix elements take a separable
form the  QRPA energies can be obtained as the roots of a relatively simple
secular equation~\cite{solo,BB}. In the case of the Skyrme interaction
this feature has been exploited by different authors \cite{gsv98,s99,n02}.

The main step of the FRSA is to simplify the central p-h interaction
$V^{C}_{ph}$ by approximating it by its Landau-Migdal form. All Landau parameters
with $l > 1$ are equal to zero in the case of Skyrme interactions. We keep only the $l=0$
terms in $V^{C}_{ph}$ and this approximation is
reasonable~\cite{gsv98,svg08}. The two-body Coulomb residual interaction
is dropped. Therefore we can write $V^{C}_{ph}$ as
\begin{eqnarray}
V^{a}_{res}({\bf r}_1,{\bf r}_2)=N_0^{-1}\left[
F_0^{a}(r_1)+G_0^{a}(r_1) {\bf \sigma}_1\cdot{\bf \sigma}_2
\right.\nonumber\\\left.+(F_0^{'a}(r_1)+G_0^{'a}(r_1){\bf \sigma
}_1\cdot{\bf \sigma}_2){\bf \tau }_1\cdot{\bf \tau }_2\right]
\delta ({\bf r}_1-{\bf r }_2), \label{res.int}
\end{eqnarray}
where $a$ is the channel index $a=\{ph,pp\}$; ${\bf \sigma}_i$ and
${\bf \tau}_i$ are the spin and isospin operators,
$N_0 = 2k_Fm^{*}/\pi^2\hbar^2$ with the Fermi momentum $k_F$ and $m^{*}$ is the
nucleon effective mass. The coefficients $G^{pp}_0$ and $G^{'pp}_0$ are zero, while the
expressions for $F^{ph}_0$, $F^{'ph}_0$, $G^{ph}_0$, $G^{'ph}_0$ and
$F^{pp}_0$, $F^{'pp}_0$ can be found in Ref.~\cite{sg81} and in
Ref.~\cite{svg08}, respectively. For the case of electric
excitations one can neglect the spin-spin terms since they play a
minor role. Though it is well known that the p-p interaction in
the spin-isospin channel ($T=0$ pairing) suppresses the $\beta^{-}$-decay half-lives,
in the present study we assume $G^{'pp}_0$=0 in order to separate
the sole impact of the tensor force. As proposed in Refs.~\cite{bzzxsc09,ss13},
we simplify the tensor p-h interaction by replacing it by the two-term
separable interaction, where the strength parameters are adjusted to
reproduce the centroid energies of the G-T and spin-quadrupole
strength distributions calculated with the original tensor p-h
interaction.

The p-h matrix elements and the antisymmetrized p-p
matrix elements can be written in a separable form in the
space of the angular coordinates~\cite{gsv98,svg08}. After integrating
over the angular variables the one-dimensional radial integrals are
numerically calculated by choosing a large enough cutoff radius
$R$ and using an $N$-point integration Gauss formula with
abscissas ${r_k}$ and weights ${w_k}$~\cite{gsv98}. Thus, one is
led to deal with a problem where the matrix elements of the residual
interaction are  sums of products and the number $\tilde N$ of terms
in the sums depends only on $N$. In particular, $\tilde N=4N+4$ and
$\tilde N=6N$ are obtained for the cases of G-T and electric excitations,
respectively. One can call it a separable approximation of finite
rank $\tilde N$ since finding the roots of the secular equation
amounts to find the zeros of a $\tilde N \times \tilde N$
determinant, and the dimensions of the determinant are independent
of the size of the configuration space, i.e., of the nucleus
considered. The studies of Refs. \cite{svg08,svg12} enable us to conclude
that $N$=45 is enough for the electric and charge-exchange
excitations considered here in nuclei with $A\le 208$.

In the next step, we construct the wave functions from a linear
combination of one-phonon and two-phonon configurations
\begin{eqnarray}
\Psi _\nu (J M) = \left(\sum_iR_i(J \nu )Q_{J M i}^{+}\right.
\nonumber\\
\left.+\sum_{\lambda _1i_1\lambda _2i_2}P_{\lambda _2i_2}^{\lambda
_1i_1}( J \nu )\left[ Q_{\lambda _1\mu _1i_1}^{+}\bar{Q}_{\lambda
_2\mu _2i_2}^{+}\right] _{J M }\right)|0\rangle~, \label{wf}
\end{eqnarray}
where $Q_{\lambda\mu i}^{+} |0\rangle$ ($\bar{Q}_{\lambda\mu i}^{+}
|0\rangle$) is the G-T (electric) excitation having energy
$\omega_{\lambda i}$ ($\bar{\omega}_{\lambda i}$).
The normalization condition for the wave functions~(\ref{wf}) is
\begin{equation}
\sum\limits_iR_i^2( J \nu)+ \sum_{\lambda _1i_1 \lambda _2i_2}
(P_{\lambda _2i_2}^{\lambda _1i_1}(J \nu))^2=1.
\end{equation}
The amplitudes $R_i(J \nu)$ and $P_{\lambda_2i_2}^{\lambda_1i_1}(J
\nu)$ are determined from the variational principle which leads to
a set of linear equations
\begin{eqnarray}
(\omega_{\lambda i}-\Omega_\nu )R_i(J \nu ) +\sum_{\lambda _1i_1
\lambda_2i_2} U_{\lambda _2i_2}^{\lambda _1i_1}(J i)
P_{\lambda_2i_2}^{\lambda _1i_1}(J \nu )=0, \label{2pheq1}
\end{eqnarray}
\begin{eqnarray}
(\omega _{\lambda _1i_1}+\bar{\omega}_{\lambda _2i_2}-\Omega_\nu
)P_{\lambda _2i_2}^{\lambda _1i_1}(J \nu)\nonumber\\
+\sum\limits_i U_{\lambda _2i_2}^{\lambda _1i_1}(J i)R_i(J \nu
)=0. \label{2pheq2}
\end{eqnarray}
The rank of the set of linear equations (\ref{2pheq1}) and
(\ref{2pheq2}) is equal to the number of one- and two-phonon
configurations included in the wave function (\ref{wf}). For its
solution it is required to compute the Hamiltonian matrix elements coupling one- and
two-phonon configurations
\begin{equation}
U_{\lambda _2i_2}^{\lambda _1i_1}(J i)= \langle 0| Q_{J i }
H \left[ Q_{\lambda _1i_1}^{+}\bar{Q}_{\lambda _2i_2}^{+}\right]
_{J} |0 \rangle.
\end{equation}
Eqs.~(\ref{2pheq1}) and (\ref{2pheq2}) have the same form as the QPM
equations~\cite{ks84,solo}, where the single-particle spectrum and
the residual interaction are derived from the same Skyrme EDF.

In the allowed G-T approximation, the $\beta^{-}$-decay rate is expressed
by summing the probabilities of the energetically allowed G-T transitions
(in units of $G_{A}^{2}/4\pi$) weighted with the integrated Fermi function
\begin{eqnarray}
T_{1/2}^{-1}=\sum_m \lambda^{m}_{if}
= D^{-1}\left(\frac{G_{A}}{G_{V}}\right)^{2}\times\nonumber\\
\sum_{m}f_{0}(Z,A,E_i-E_{1_m^+})B(G-T)_{m},
\label{t}
\end{eqnarray}
\begin{equation}
E_i-E_{1_m^+}\approx\Delta M_{n-H}+\mu_n-\mu_p-E_m, \label{et}
\end{equation}
where $\lambda^{m}_{if}$ is the partial $\beta^{-}$-decay rate,
$G_A/G_V$=1.25~\cite{Suhonen} is
the ratio of the weak axial-vector and vector coupling
constants  and $D$=6147 s
(see Ref.~\cite{Suhonen}). $\Delta M_{n-H}=0.782$~MeV is the mass
difference between the neutron and the hydrogen atom,  $\mu_n$ and
$\mu_p$ are the neutron and proton chemical potentials,
respectively, $E_i$ is the ground state energy of the parent
nucleus, and $E_{1_m^+}$ denotes a state of the daughter nucleus
$(Z,A)$. $E_m$ and $B(G-T)_{m}$ are the solutions either of the
QRPA equations, or of Eqs.~(\ref{2pheq1})-(\ref{2pheq2}) taking
into account the two-phonon configurations.

Thus, to calculate the half-lives by Eqs.(\ref{t}) and (\ref{et}) an approximation
worked out in Ref.~\cite{ebnds99} is used. It allows one to avoid an implicit calculation
of the nuclear masses and $Q_{\beta}$-values. However, one should realize that
the related uncertainty in constraining the parent nucleus ground state calculated
with the chosen Skyrme interaction is transferred to the values
of the neutron and proton chemical potentials entering Eq.~(\ref{et}).
%
%==============================================================
%
\section{Details of calculations}
\begin{figure}[t!]
\includegraphics[width=1.0\columnwidth]{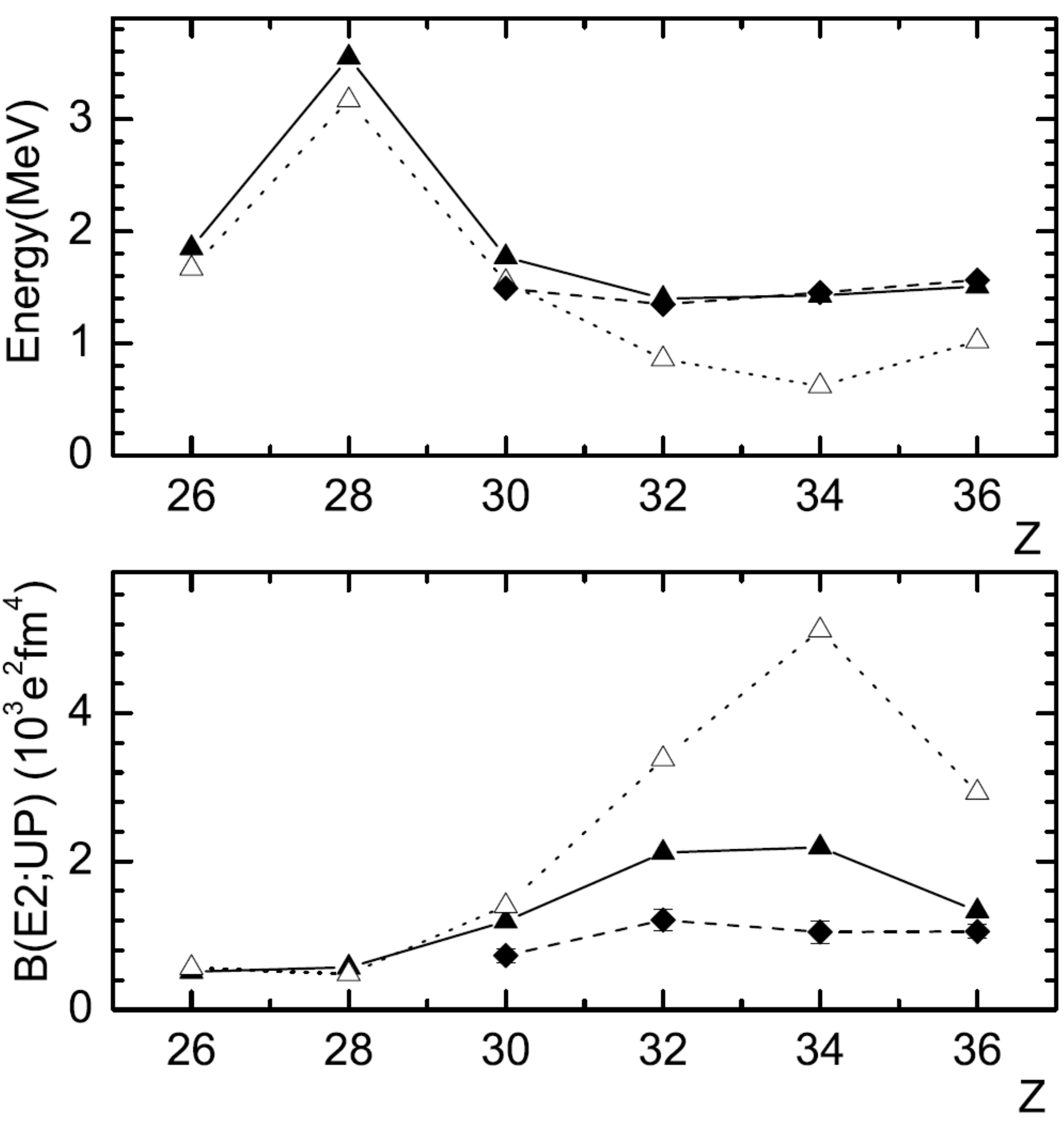}
\caption{Energies and $B(E2)$ values for up-transitions to the
$[2_1^+]_{QRPA}$ states in the neutron-rich $N=50$ isotones.
Results of the calculations without the tensor interaction (open
triangles) and with the tensor interaction (filled triangles) are
shown. Experimental data (filled diamonds) are taken from
Ref.~\cite{pbsh13}.}
\end{figure}
\begin{figure}[t!]
\includegraphics[width=1.0\columnwidth]{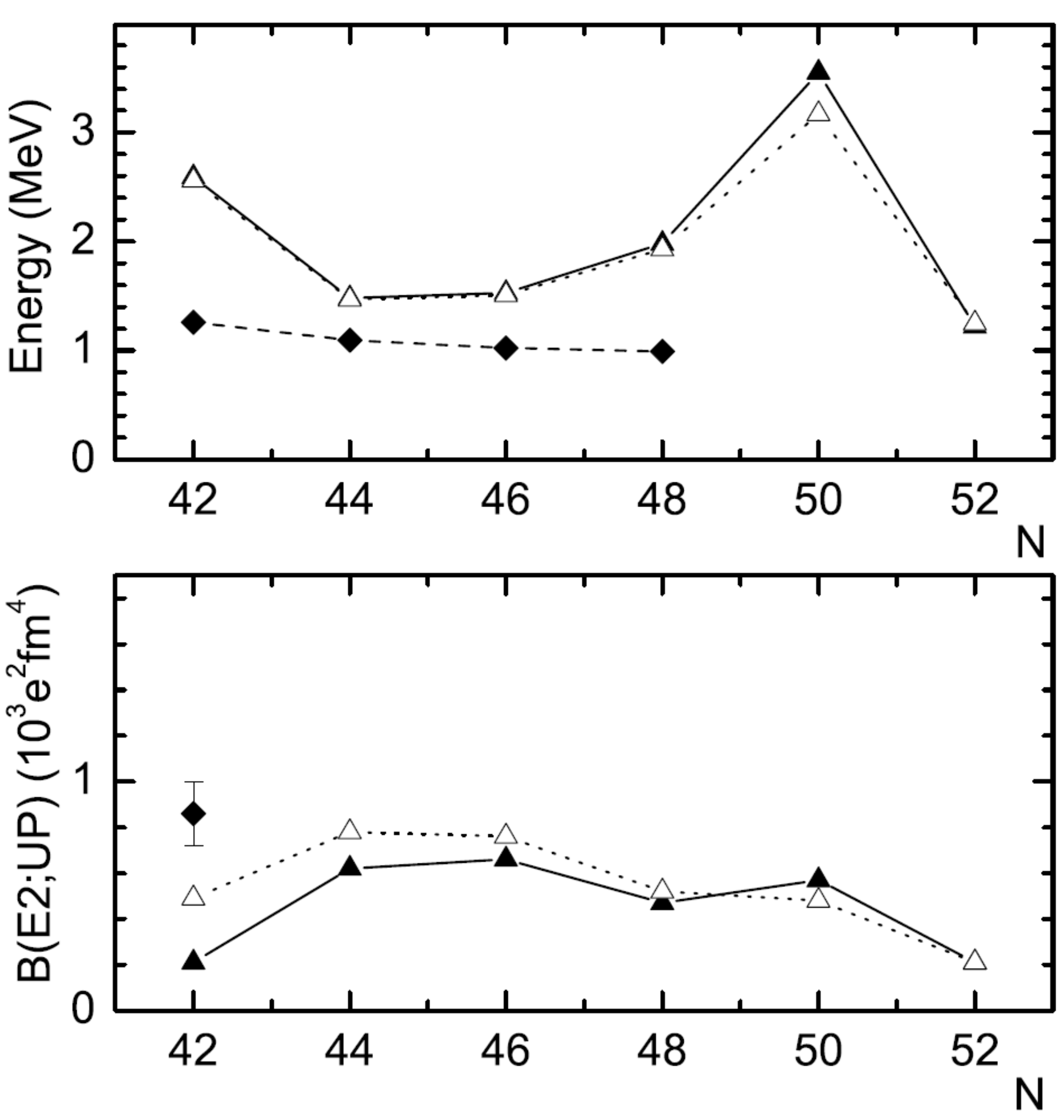}
\caption{Same as Fig. 1, but for the neutron-rich Ni isotopes}
\end{figure}
We apply the approach to study the influence of the coupling
between one- and two-phonon terms in the wave functions, as well as the
tensor force effects on the strength distributions of G-T states
in the neutron-rich Ni isotopes and $N=50$ isotones.
To obtain the interaction in the p-h channel, we use the Skyrme interaction
SGII~\cite{sg81} and the zero-range tensor interaction of Ref.~\cite{bzzxsc09}
with $\alpha_T$=-180~MeVfm$^5$ and $\beta_T$=120~MeVfm$^5$ for the parameters
$\alpha$ and $\beta$ of the one-body spin-orbit potential of Eq.(\ref{sop}).

Since the SGII parametrization gives
reasonable values for the Landau parameters $F^{'}_{0}=0.73$
and $G^{'}_{0}=0.50$, one obtains a successful description of the
spin-dependent properties, in particular for the experimental
energies of the G-T resonances of $^{90}$Zr~\cite{sg81}. In
Ref.~\cite{ss13}, the FRSA model has been used
to calculate the G-T states of $^{90}$Zr and $^{208}$Pb, and the FRSA
results reproduce the main features of the G-T strength
distributions obtained with the exact treatment of Skyrme tensor
interactions in the RPA.

For the interaction in the p-p channel, we use a zero-range
volume force, i.e., $\eta=$0 in Eq.~(\ref{pair}) and
$V_0$=-270 MeVfm$^{3}$ with a smooth cut-off at 10 MeV above the
Fermi energies~\cite{svg08,k90}. This value of the pairing
strength has been fitted to reproduce the experimental pairing
energies of $^{70,72,74,76}$Ni obtained from binding energies of
neighbouring nuclei. This choice of the pairing interaction has
also been used for a satisfactory description of the experimental
data of $^{90,92}$Zr and $^{92,94}$Mo~\cite{sap12}. Because of the
closed $Z=28$ and $N=50$ shells, the $T=0$ pairing is not effective
in these nuclei~\cite{nmvpr05} and, therefore, it can be neglected.

We now use the FRSA of the residual interaction and carry out
QRPA calculations in very large two-quasiparticle spaces.
In particular, the cut-off in the discretized continuous part of
the s.p. spectra is at 100~MeV. Because of the inclusion of the
tensor correlation effects within the $1p-1h$ and $2p-2h$
configuration space, we do not need any quenching
factor~\cite{bh82}. For the ansatz of the wave function
(\ref{wf}), the Ikeda sum rule $S_{-}-S_{+}=3(N-Z)$ is fulfilled, see e.g.,
\cite{ks84}. Our configurational space is sufficient to exhaust
this sum rule for the G-T strength of the nuclei that we studied
without and with the tensor force.  As an illustrative example, in
the case taking into account the tensor correlations for
$^{82}_{32}$Ge we obtain $S_{-}=56.85$ and $S_{+}=2.84$ in the QRPA,
the inclusion of the 2p-2h effects gives the same values. The results
without the tensor interaction indicate $S_{-}=54.46$ and $S_{+}=0.45$.

Let us now focus on the properties of the low-energy G-T state,
since we are interested in the $\beta$-decay half-lives.
Its experimentally known half-life puts an indirect constrain on the
calculated G-T strength distributions within the $Q_{\beta}$-window.
Let us examine the extension of the configuration space to one- and
two-phonon terms. To construct the wave functions~(\ref{wf}) of the
low-lying $1^{+}$ states we use only the
$[1^{+}_{i}\otimes\lambda^{+}_{i'}]_{QRPA}$ terms
and all electric phonons with $\lambda > 2$ vanish, i.e., G-T phonons from
the charge-exchange modes are only used in the two-phonon terms,
as in Ref.~\cite{cnbb94,ncbbm12}. All one- and two-phonon configurations
with the transition energies  $|E_{1_m^+}-E_i|$ up to 10~MeV are included.
We have checked that the inclusion of the high-energy configurations
leads to minor effects on the half-life values.

It is interesting to study the energies, reduced transition
probabilities and the structure of the $2_1^+$ QRPA state. The
calculated $2_1^+$ energies and transition probabilities in the
neutron-rich $N=50$ isotones are compared with existing
experimental data~\cite{pbsh13} in Fig.1. The FRSA model with the
tensor interaction reproduces the experimental data~\cite{pbsh13}
very well. We find that the tensor interaction induces a reduction
of the $2_1^+$ collectivity and it results in a decrease
of the transition probability, see Fig.1. There is a remarkable
increase of the $2_1^+$ energy of $^{78}$Ni in comparison with
those in $^{76}$Fe and $^{80}$Zn. It corresponds to a standard
evolution of the $2_1^+$ energy near closed shells. As can be seen
from Fig.2, the behavior of $2_1^+$ energies of $^{76,78,80}$Ni is
similar to that of the $N=50$ isotones. It is seen that the
inclusion of the tensor interaction does not change energies and
transition probabilities along this Ni isotopic chain except for
$^{70}$Ni. Including the tensor interaction changes contributions
of the main configurations only slightly, but the general
structure of the $2_1^+$ state remains the same. The neutron
amplitudes are dominant in all Ni isotopes and the contribution of
the main neutron configuration $\{1g_{9/2},1g_{9/2}\}$ decreases
from 89\% for $^{70}$Ni to 77\% for $^{76}$Ni when neutrons fill
the subshell $1g_{9/2}$. Thus, a satisfactory description of
$2_1^+$ energies is found for $^{72,74,76}$Ni. Our calculated
$2_1^+$ energy of $^{70}$Ni is about a factor of 2 too large
compared to the data value~\cite{pbsh13}. This is likely due to the
overestimation of the neutron contribution to the $2_1^+$ wave
function within the QRPA. One can expect some improvement if the
coupling with the two-phonon components of the wave functions is
taken into account~\cite{svg04,sap12}.
\begin{figure}[t!]
\includegraphics[width=1.0\columnwidth]{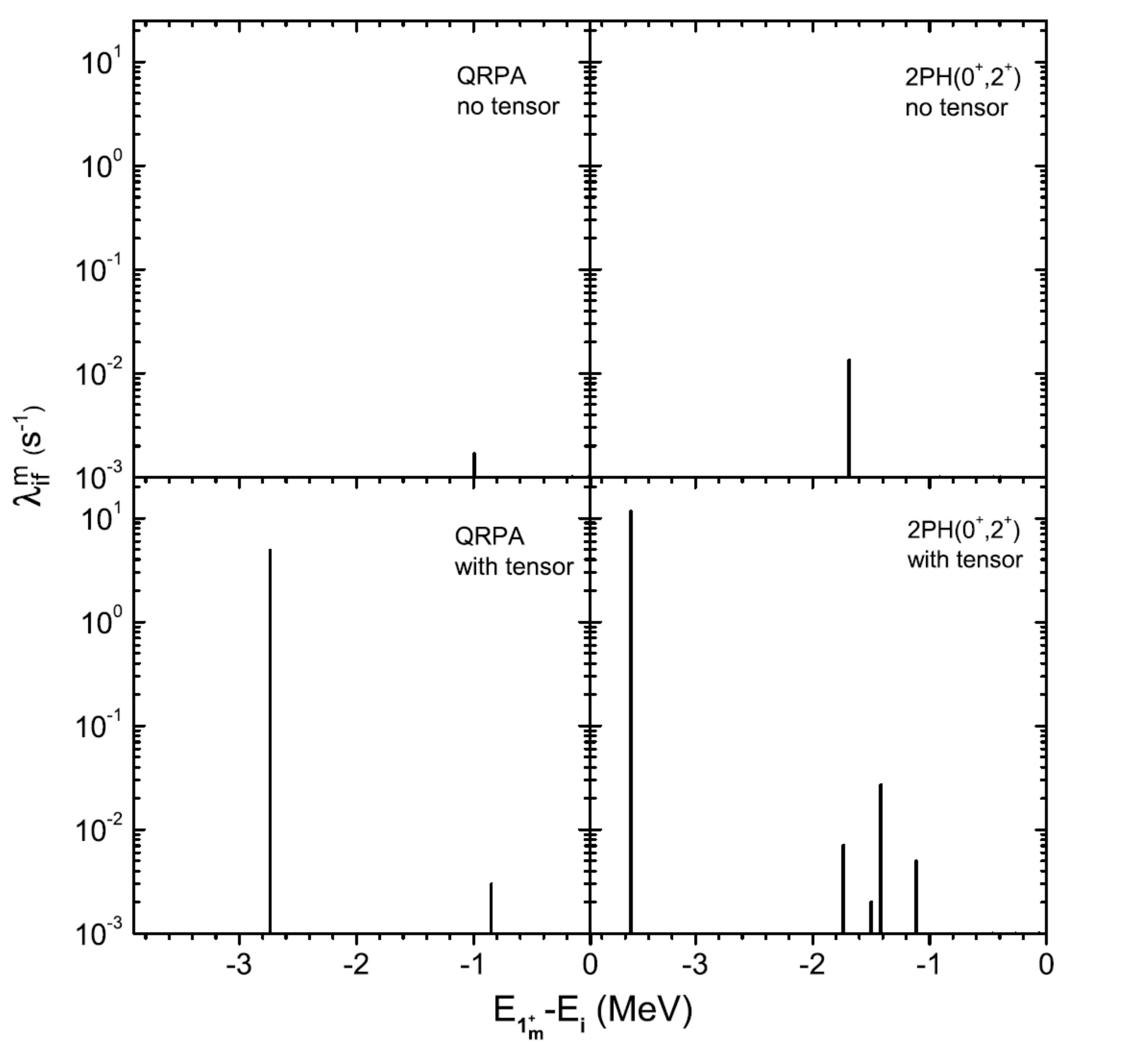}
\caption{The phonon-phonon coupling effect on the
$\beta$-transition rates in $^{80}$Zn. The left and right panels correspond
to the calculations within the  QRPA and taking into account
the $2p-2h$ configurations, respectively. Results of the calculations
without (resp. with) the tensor interaction are shown in the upper (resp. lower) panels.}
\end{figure}
\begin{figure}[t!]
\includegraphics[width=1.0\columnwidth]{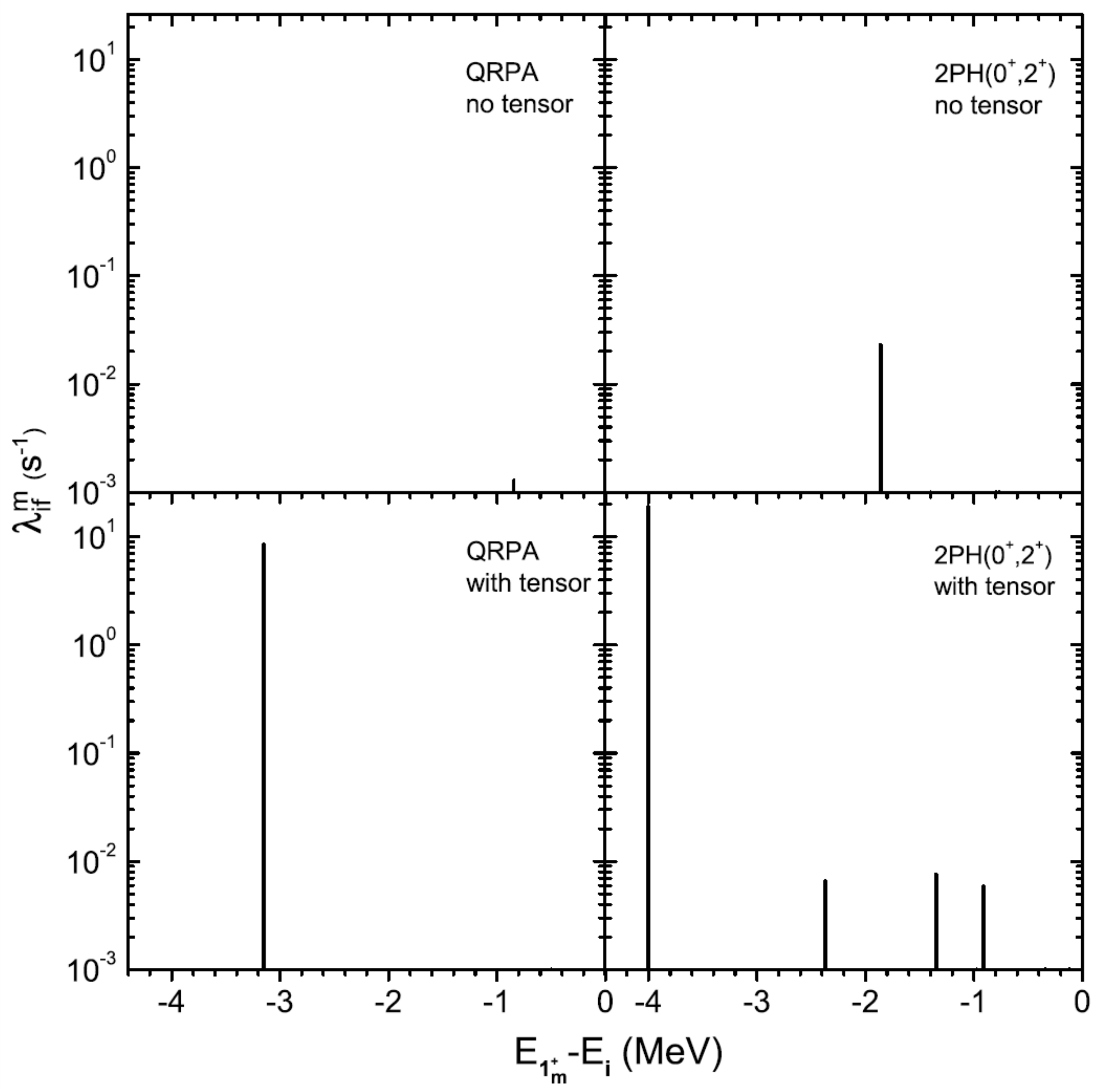}
\caption{Same as Fig. 3, but for $^{72}$Ni.}
\end{figure}
\begin{figure}[t!]
\includegraphics[width=1.0\columnwidth]{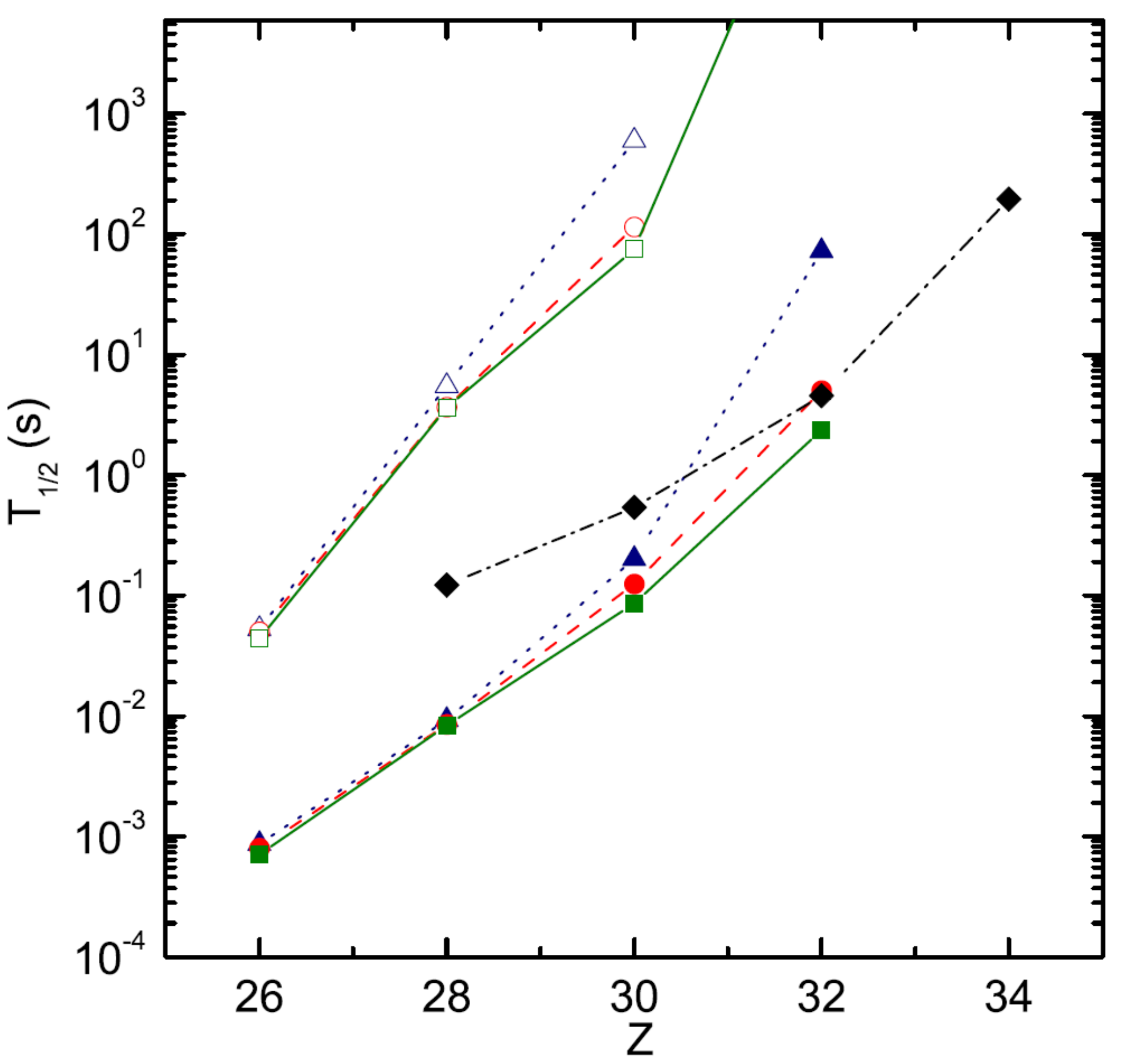}
\caption{The phonon-phonon coupling effect on $\beta^-$-decay
half-lives of the neutron-rich $N=50$ isotones.
Results of the calculations without the tensor interaction (open triangles, circles, squares)
and with the tensor interaction (filled triangles, circles, squares) are shown. The squares
correspond to the half-lives calculated with inclusion of the phonon-phonon coupling,
the triangles are the QRPA calculations. Results including the
$[1^{+}_{i}\otimes 2^{+}_{i'}]_{QRPA}$ configurations are denoted by the circles.
Experimental data (filled diamonds) are from Refs.~\cite{expthalflives,Xu}.}
\end{figure}
\begin{figure}[t!]
\includegraphics[width=1.0\columnwidth]{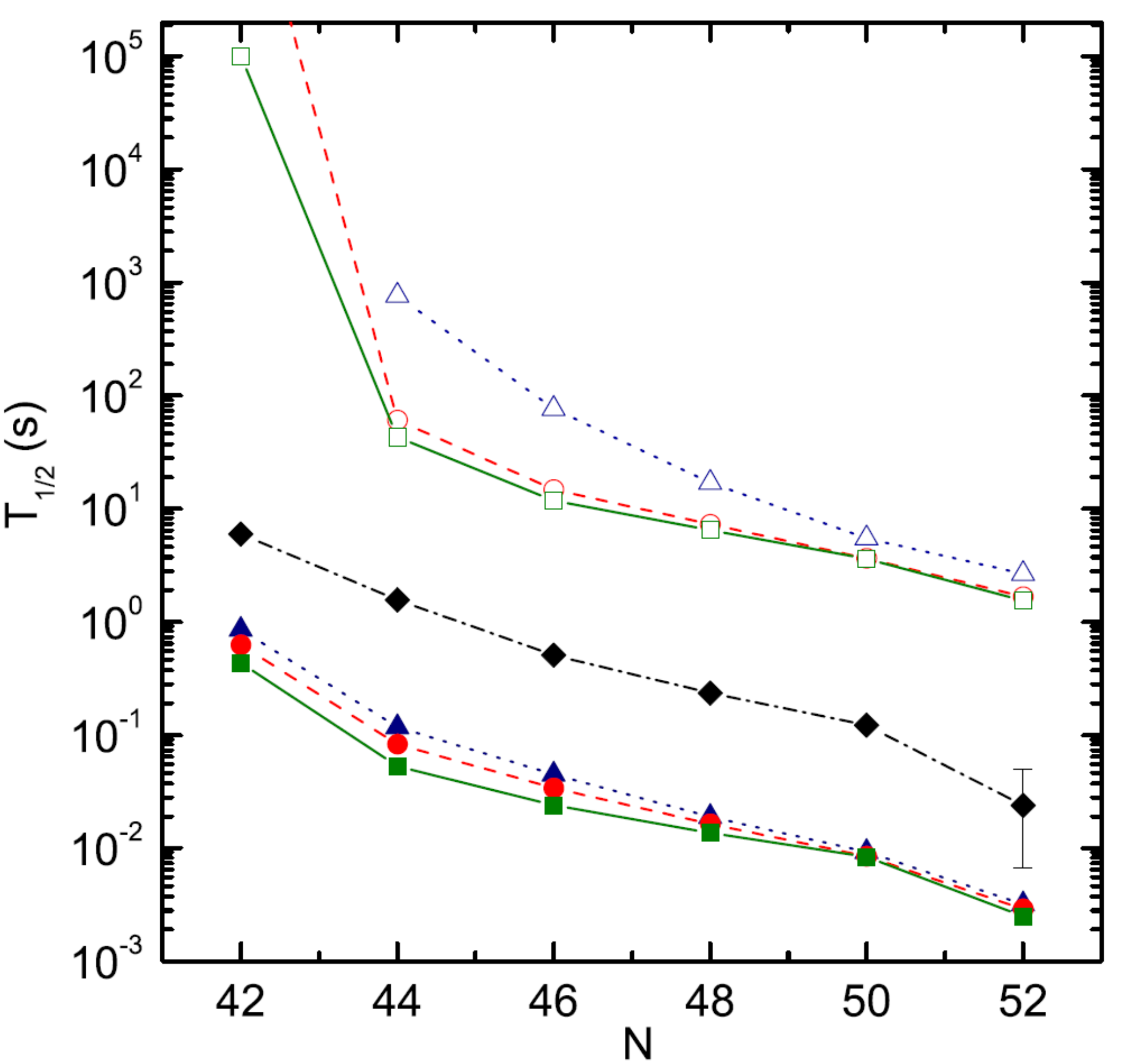}
\caption{Same as Fig. 5, but for the neutron-rich Ni isotopes.
Experimental data are taken from Refs.~\cite{expthalflives,Xu,6874ni}.}
\end{figure}
%
%
%==============================================================
%
\section{Results for $\beta$-decay rates}
Using the same set of parameters we calculate the low-lying
G-T strength distributions of the neutron-rich nuclei
$^{76}_{26}$Fe,
$^{70,72,74,76,78,80}_{\,\,\,\,\,\,\,\,\,\,\,\,\,\,\,\,\,\,\,\,\,\,\,\,\,\,\,\,\,\,\,\,\,\,\,\,\,\,28}$Ni,
$^{80}_{30}$Zn, $^{82}_{32}$Ge and $^{84}_{34}$Se.
First, the properties of the low-lying $1^+$ states of the daughter nuclei
are studied within the one-phonon approximation.  As expected, the largest
contribution ($>$88\%) in the calculated $\beta^-$-decay half-life
comes from the $[1_1^+]_{QRPA}$ state. To illustrate it,
the $\beta$-transition rates $\lambda^{m}_{if}$ of $^{80}$Zn
and $^{72}$Ni are shown in Fig.~3 and Fig.~4, respectively.
The transition energies $E_{1^+_m}-E_i$ refer to the ground state of
the parent nucleus.

QRPA results calculated without the tensor interaction indicate
that the dominant configuration of the $[1_1^+]_{QRPA}$ states is
$\{\pi2p\frac{3}{2}\nu2p\frac{1}{2}\}$ whose contribution is about
91~$\%$ in all the cases considered. In other words, the
$[1_1^+]_{QRPA}$ state is not collective and, therefore, the
$\beta^-$-decay is related to the unperturbed
$\{\pi2p\frac{3}{2}\nu2p\frac{1}{2}\}$ energy. At the same time
the calculated $\log ft$ value increases from 3.7 for
$^{72}$Ni to 3.9 for $^{80}$Ni. The $\beta^-$-decay
half-lives of the neutron-rich $N=50$ isotones and the Ni isotopes
are shown in Fig.~5 and Fig.~6, respectively. The calculated
values overestimate the experimental
data~\cite{78ni,6874ni,expthalflives,Xu}. Moreover, for $^{70}$Ni,
$^{82}$Ge and $^{84}$Se, the $[1_1^+]_{QRPA}$ state of the
daughter nucleus is above the parent ground state, i.e.,
this calculation predicts stable $^{70}$Ni, $^{82}$Ge and $^{84}$Se.
It is worth pointing out that the relativistic QRPA
approach~\cite{nmvpr05} gives a similar structure of the $[1_1^+]_{QRPA}$
state for these nuclei.

As a general trend, we observe a redistribution of the G-T strengths by the
inclusion of the tensor interaction in the QRPA, in the same way as it
was found in Refs.~\cite{bszzcx09,sag11,ss13}. The tensor correlations
shift up about 10\% of the G-T strength to the energy region above 30MeV.
Also, the tensor interaction makes a downward shift of
the strength in the G-T resonance region below the peak energy and
the $[1_1^+]_{QRPA}$ state is moved downwards. This is illustrated in
the cases of $^{80}$Zn and $^{72}$Ni (see  left bottom panels of
Figs.~3-4). For $^{80}$Ni, the $[1_1^+]_{QRPA}$ energy shift reaches 2.7~MeV.
In particular, because of this shift, we get unstable $^{70}$Ni and $^{82}$Ge.
In addition, the tensor correlations lead to a collective structure for the
$[1_1^+]_{QRPA}$ state with the dominance of the
$\{\pi2p\frac{3}{2}\nu2p\frac{1}{2}\}$ configuration. In the case
of the $N=50$ isotones the $\{\pi2p\frac{3}{2}\nu2p\frac{1}{2}\}$
contribution increases from 46\% in $^{78}$Ni to 72\% in $^{84}$Se, whereas
for Ni isotopes this contribution decreases from 57\% in $^{70}$Ni to 39\%
in $^{80}$Ni. The $[1_1^+]_{QRPA}$ collectivity is reflected in the $\log ft$
value and there is a slight decrease of the $\log ft$ from 2.3 for $^{70}$Ni
to 2.0 for $^{80}$Ni. As can be seen from  Fig.5 and Fig.6, the half-lives
calculated with the tensor force are about 60-2000 times shorter than
those calculated without.
This analysis within the one-phonon approximation
can help to identify the tensor correlation effects, but it is only
a rough estimate. It is worth to mention that the first discussion
of the strong impact of the tensor correlations on the $\beta^-$-decay
half-lives based on QRPA calculations with Skyrme forces has
been done in Ref.~\cite{mb13}.

Let us now discuss the extension of the space to one- and
two-phonon configurations in the FRSA model when the tensor
interaction is taken into account. In all ten nuclei, the dominant
contribution in the wave function of the first $1^+$ state comes
from the $[1^+_1]_{QRPA}$ configuration, but the two-phonon
contributions are appreciable. The main two-phonon components of
the $1^+_1$ wave function are the $[1_1^+\otimes2_1^+]_{QRPA}$ and
$[1_1^+\otimes0_2^+]_{QRPA}$ configurations. When the tensor
interaction is not included, the main configuration is only the
$[1_1^+\otimes2_1^+]_{QRPA}$ configuration. As a result, the
inclusion of the two-phonon terms results in an increase of the
transition energies $|E_{1_1^+}-E_i|$ and the energy shift is
large (1.1 MeV  for $^{82}$Ge and 0.8 MeV  for $^{80}$Zn) in
comparison with 0.2 MeV in the case of the doubly-magic nucleus
$^{78}$Ni. The $\log ft$ value is practically unchanged. One can
see from Figs.5-6 that the effects of the phonon-phonon coupling
produce a sizable impact on the $\beta$-decay half-life which is
reduced by a factor 30. Specifically, the
$[1_1^+\otimes2_1^+]_{QRPA}$ configuration is the important
ingredient for the half-life description since the
$[2_1^+]_{QRPA}$ state is the lowest collective-electromagnetic
excitation which leads to the minimal two-phonon energy and the
maximal matrix elements coupling one- and two-phonon
configurations of Eqs.~(\ref{2pheq1}) and (\ref{2pheq2}). Figs. 5
and 6 also show the half-life reduction as an effect of the
quadrupole-phonon coupling. For $^{84}$Se, the experimental
half-life is $T_{1/2}=3.26\pm0.10$~min\cite{expthalflives}, but
our model predicts that this nucleus is stable against
$\beta$-decay and, in particular, $E_{1_1^+}-E_i$=0.0~MeV. One can
expect an improvement if the $T=0$ pairing interaction is taken
into account. We intend to extend our formalism to include the $T=0$
pairing effects.

It is interesting to discuss the change of the half-lives along
the chains of the $N=50$ isotones and the Ni isotopes. As pointed
out in Refs.~\cite{expthalflives,6874ni,Xu}, there are different
evolutions of the existing experimental half-lives, namely, the 37.3-time
reduction of half-life values from $^{82}$Ge to $^{78}$Ni and the
gradual reduction of half-lives with increasing neutron number for
the Ni isotopes. One can see that our results reproduce this behaviour
which is sensitive to the isotopic and isotonic dependences of the
transition energies $|E_{1_1^+}-E_i|$.  We find that the energy $|E_{1_1^+}-E_i|$
for $^{82}$Ge ($^{74}$Ni) is 3.1 (1.2) times less than that
in $^{78}$Ni. It is worth to mention that the calculated half-life of
$^{82}$Ge is in reasonable agreement with the
experimental data.  As can be seen from Fig. 6, the main discrepancies
between measured and calculated half-lives of Ni isotopes are due to
too strong tensor correlations and one should seek for improvements
in the tensor part of the effective interaction used. Finally, for
neighbours of $^{78}$Ni we give predictions as the bottom limit of
the half-life (0.7 ms for $^{76}$Fe and 2.5 ms for $^{80}$Ni).
It is seen that the phonon-phonon coupling plays a minor role.
Our calculated half-life of $^{80}$Ni is in qualitative agreement with
the recently observed value of $23.9^{+26.0}_{-17.2}$ms~\cite{Xu}.

The SGII parametrization of the central force gives a
reasonable description of properties of the G-T and
charge-exchange spin-dipole resonances of
$^{90}$Zr~\cite{sag11,sg81} and the low-energy spectrum of the
quadrupole excitations for nuclei near $^{90}$Zr~\cite{sap12}. For
the half-life description, the quantitative agreement with the
experimental data is not satisfactory for the neutron-rich nuclei.
A possible reason might be the underestimated symmetry energy
of 26.8~MeV in the case of the SGII set. A half-life study of the
influence of the two-phonon terms with taking into account the
tensor interaction  for  the different parametrizations of central
Skyrme force is still underway.

Inclusion of the tensor interaction results in significant increase of
the transition energies and partial rates of the main G-T transitions.
Including the $2p-2h$ configurations leads to further increase of its
rate and to the appearance of the weak fragmented satellites at low transition
energy (see, e.g., right bottom panels of Figs.3-4). Importantly, such
a change of the G-T strength distribution takes place in the
near-threshold region. Thus, an additional constraint on the $\beta$-strength distribution
is also given by delayed neutron emission probability
($P_n$-value)~\cite{pn05}. Since the FRSA model enables one to
evaluate the coupling of QRPA phonons to more complex
configurations, such calculations that take into account the $2p-2h$
fragmentation of the QRPA excitations are now in progress.
%
%=====================================================================
%
\section{Conclusions}
Starting from a Skyrme effective interaction the G-T strength in the $Q_{\beta}$- window
has been studied within the extended FRSA model including both the tensor
interaction and $2p-2h$ configurations effects. The suggested
approach enables one to perform the calculations in very large
configuration spaces. Using the parameter set SGII+tensor
interaction, we have applied this model to the G-T states in the
neutron-rich Ni isotopes and $N=50$ isotones, for which
experimental data of $\beta^{-}$-decay rates are available.
The inclusion of the tensor interaction leads to a redistribution
of the G-T strength. The low-energy G-T strength
is fragmented and the $1^+_1$ state is moved downwards.
We observe that the coupling between one- and two-phonon terms
with the $2^+$ phonon states is strong in these nuclei, and
the $2p-2h$ fragmentation  and damping of the QRPA excitations
are thus important. In particular, the energy  shift of
the lowest $1^+$ state due to the phonon-phonon coupling is large:
0.8 MeV in $^{80}$Zn and 1.1 MeV in $^{82}$Ge in comparison with
0.2 MeV in the case of $^{78}$Ni. Taking into account these effects
results in a dramatic reduction of $\beta$-decay half-lives.
It is shown that the $2p-2h$ impact on the half-lives comes inherently
from the $[1_1^+\otimes2_1^+]_{QRPA}$ term of the wave function of
the $1^+_1$ state. At a qualitative level, our results reproduce
the experimental evolution of the half-lives, i.e., for the $N=50$
isotones we describe the sharp reduction of half-lives with decreasing
proton number and at the same time, for the Ni isotopes, the gradual
reduction of half-lives with increasing neutron number.
We give predictions for open-shell nuclei $^{76}$Fe and $^{80}$Ni
near $^{78}$Ni that are important for stellar nucleosynthesis.
Using the strong tensor correlations~\cite{bzzxsc09} our estimation
is rather the bottom limit of these half-lives. Also, our calculations
show that the influence of the phonon coupling is small on the
half-lives of $^{76}$Fe and $^{80}$Ni.

More information on the $\beta$-strength distributions can be derived
from simultaneous analysis of the half-lives and the $P_n$ values~\cite{pn05,Miernik}.
Such a program of detailed calculations  is underway, the results will be
reported in connection to the RIKEN experiments on $^{78}$Ni~\cite{RIKEN}.
%
%==================================================================
%
\section*{Acknowledgments}
A.P.S., V.V.V., I.N.B and N.N.A thank the hospitality of IPN-Orsay where
the part of this work was done. This work is partly supported by
the IN2P3-RFBR agreement No.~110291054 and the IN2P3-JINR agreement.
%
%==================================================================
%

%
%==================================================================
%
\end{document}